\documentclass[12pt]{article}

\usepackage{amsmath}
\usepackage{amsthm}
\usepackage{titlesec}
\usepackage{indentfirst}
\usepackage{amssymb}
\usepackage{graphics}
\usepackage{latexsym}
\usepackage{cite}

\theoremstyle{definition}\newtheorem{Df}{Definition}
\theoremstyle{plain}\newtheorem{Th}{Theorem}
\theoremstyle{definition}\newtheorem{Rm}{Remark}
\theoremstyle{definition}\newtheorem{Emp}{Example}
\theoremstyle{plain}\newtheorem{Pp}[Th]{Proposition}
\theoremstyle{plain}
\theoremstyle{plain}\newtheorem{Lm}[Th]{Lemma} \textwidth 175mm
\allowdisplaybreaks

\begin{document}
 \title{{\bf  Minimum-error discrimination of quantum states: New bounds and comparison}\thanks
 {This work is supported by the National Natural Science Foundation
(Nos. 60573006, 60873055), the Research Foundation for the
Doctoral Program of Higher School of Ministry of Education (No.
20050558015), and NCET of China.}}

\author{Daowen Qiu$^{a,b,}$\thanks{
{\it E-mail address:} issqdw@mail.sysu.edu.cn (D.W. Qiu).} ,\hskip
3mm Lvjun Li$^{a,} $\thanks{
{\it E-mail address:} hnbcllj@163.com (L.J. Li).} \\
\small{{\it $^a$Department of
Computer Science, Zhongshan University, Guangzhou 510275, China}}\\
\small{{\it $^b$SQIG--Instituto de Telecomunica\c{c}\~{o}es, IST,
TULisbon, }}\\
\small{{\it  Av. Rovisco Pais 1049-001, Lisbon, Portugal}}\\
}
\date{ }
\maketitle

\date{ }
\maketitle
\begin{center}
\begin{minipage}{130mm}
\begin{center}{\bf Abstract}\end{center}
{\small

The minimum-error probability of ambiguous discrimination for two
quantum states is the well-known {\it Helstrom limit} presented in
1976. Since then, it has been thought of as an intractable problem
to obtain the minimum-error probability for ambiguously
discriminating arbitrary $m$ quantum states. In this paper, we
obtain a new lower bound on the minimum-error probability for
ambiguous discrimination and compare this bound with six other
bounds in the literature. Moreover, we show that the bound between
ambiguous and unambiguous discrimination does not extend to
ensembles of more than two states. Specifically, the main
technical contributions are described as follows: (1) We derive a
new lower bound on the minimum-error probability for ambiguous
discrimination among arbitrary $m$ mixed quantum states with given
prior probabilities, and we present a necessary and sufficient
condition to show that this lower bound is attainable. (2) We
compare this new lower bound with six other  bounds in the
literature in detail, and, in some cases, this bound is optimal.
(3) It is known that if $m=2$, the optimal inconclusive
probability of unambiguous discrimination $Q_{U}$ and the
minimum-error probability of ambiguous discrimination $Q_{E}$
between arbitrary given $m$ mixed quantum states have the
relationship $Q_{U}\geq 2Q_{E}$.  In this paper, we show that,
however, if $m>2$, the relationship $Q_{U}\geq 2Q_{E}$ may not
hold again in general, and there may be no supremum of $Q_{U}/Q_{E}$ for more than two states, which may also reflect an essential
difference between discrimination for two-states and multi-states. 
(4) A number of examples are constructed. }


\end{minipage}
\end{center}
\vskip 10mm

\section*{I. Introduction}

A fundamental issue in quantum information science is that
nonorthogonal quantum states cannot be perfectly discriminated,
and indeed, motivated by the study of quantum communication and
quantum cryptography \cite{GRTZ02}, distinguishing quantum states
has become a more and more  important subject in quantum
information theory
\cite{Hel76,Che00,BHH04,Che04,BC08,EF01,Hol73,YKL75}. This problem
may be roughly described by the connection between quantum
communication and quantum state discrimination in this manner
\cite{Hel76,Che00,BC08,Hol73,YKL75}: Suppose that a transmitter,
Alice, wants to convey classical information to a receiver, Bob,
using a quantum channel, and Alice represents the message conveyed
as a mixed quantum state that, with given prior probabilities,
belongs to a finite set of mixed quantum states, say
$\{\rho_{1},\rho_{2},\ldots,\rho_{m}\}$; then Bob identifies the
state by a measurement.

As it is known, if the supports of mixed states
$\rho_{1},\rho_{2},\ldots,\rho_{m}$ are not mutually orthogonal,
then Bob can not reliably identify which state Alice has sent,
namely, $\rho_{1},\rho_{2},\ldots,\rho_{m}$ can not be faithfully
distinguished  \cite{Hel76,Hol73,YKL75}. However, it is always
possible to discriminate them in a probabilistic means. To date,
there have been many interesting results concerning quantum state
discrimination, we may refer to \cite{Che00,BHH04,BC08,Qiu08} and
the references therein. It is worth mentioning that some schemes
of quantum state discrimination have been experimentally realized
(for example, see \cite{BR97,CKCBR01,MSB04} and the detailed
review in \cite{BC08}).

Various strategies have been proposed for distinguishing quantum
states. Assume that  mixed states
$\rho_{1},\rho_{2},\ldots,\rho_{m}$ have the {\it a priori}
probabilities $p_{1},p_{2},\ldots,p_{m}$, respectively. In
general, there are three fashions to discriminate them. The first
approach is {\it ambiguous discrimination} (also called {\it
quantum state detection}) \cite{Hel76,Hol73,YKL75} that will be
further studied in this paper, in which inconclusive outcome is
not allowed, and thus error may result. A measurement for
discrimination consists of $m$ measurement operators (e.g.,
positive semidefinite operators) that form a resolution of the
identity on the Hilbert space spanned by the all eigenvectors
corresponding to all nonzero eigenvalues of
$\rho_{1},\rho_{2},\ldots,\rho_{m}$. Much work has been devoted to
devising a measurement maximizing the success probability (i.e.,
minimizing the error probability) for detecting the states
\cite{CBH89,EMV03,OBH96,BKMH97,EMV04}.

The first important result is the pioneering work by Helstrom
\cite{Hel76}---a general expression of the minimum achievable
error probability for distinguishing between two mixed quantum
states. For the case of more than two quantum states, some
necessary and sufficient conditions have been derived for an
optimum measurement maximizing the success probability of correct
detection \cite{Hol73,YKL75,EMV03}. However, analytical solutions
for an optimum measurement have been obtained only for some
special cases (see, for example, \cite{Bar01,ABGH02,CH03}).

Regarding the minimum-error probability for ambiguous
discrimination between arbitrary $m$ mixed quantum states with
given prior probabilities, Hayashi et al. \cite{HKK06} gave a
lower bound  in terms of the individual operator norm.  Recently,
Qiu \cite{Qiu08} obtained a different lower bound by means of
pairwise trace distance. When $m=2$, these two bounds are
precisely the well-known Helstrom limit \cite{Hel76}. Afterwards,
Montanaro \cite{Mon08} derived another lower bound by virtue of
pairwise fidelity. However, when $m=2$, the lower bound in
\cite{Mon08} is smaller than Helstrom limit. Indeed, it is worth
mentioning that, with a lemma by Nayak and Salzman \cite{NS06}, we
can also obtain a different lower bound represented by the prior
probabilities (we will review these bounds in detail in Section
II). Besides this, there also exist the other lower bounds
\cite{M07,T09}, and upper bounds \cite{BK02,HLS05}.

The second approach is the so-called {\it unambiguous
discrimination} \cite{Che00,Iva87,Die88,Per88,JS95,PT98}, first
suggested by Ivanovic, Dicks, and Peres \cite{Iva87,Die88,Per88}
for the discrimination of two pure states. In contrast to
ambiguous discrimination, unambiguous discrimination allows an
inconclusive result to be returned, but no error occurs. In other
words,  this basic idea for distinguishing between $m$ pure states
is to devise a measurement that with a certain probability returns
an inconclusive result, but, if the measurement returns an answer,
then the answer is fully correct. Therefore, such a measurement
consists of $m+1$ measurement operators, in which a measurement
operator returns an inconclusive outcome. Analytical solutions for
the optimal failure probabilities have been given for
distinguishing between two and three pure states
\cite{Iva87,Die88,Per88,JS95,PT98,DG98}. Chefles \cite{Che98}
showed that a set of pure states is amendable to unambiguous
discrimination if and only if they are linearly independent.  The
optimal unambiguous discrimination between linearly independent
symmetric and equiprobable pure states was solved in \cite{CB98}.
A semidefinite programming approach to unambiguous discrimination
between pure states has been investigated in detail by Eldar
\cite{Eld03IT}. Some upper bounds on the optimal success
probability for unambiguous discrimination between pure states
have also been presented (see, for example,
\cite{Qiu02L,Qiu02J,Zha01,BC08} and references therein).

We briefly recollect unambiguous discrimination between  mixed
quantum states. In \cite{RST03,FDY04}, general upper and lower
bounds on the optimal failure probability for distinguishing
between two and more than two mixed quantum states have been
derived. The analytical results for the optimal unambiguous
discrimination between two mixed quantum states have been derived
in \cite{RLE03,HB05}. For more work regarding unambiguous
discrimination, we may refer to \cite{BHH04,BC08}.

The third strategy for discrimination combines the former two
methods \cite{CheB98,FJ03,Eld03}. That is to say, under the
condition that a fixed probability of inconclusive outcome is
allowed to occur, one tries to determine the minimum achievable
probability of errors for ambiguous discrimination. Such a scheme
for discriminating pure states has been considered in
\cite{CheB98,FJ03}, and, for discrimination of mixed states, it
was dealt with in \cite{Eld03}.  Indeed, by allowing for an
inconclusive result occurring, then one can obtain a higher
probability of correct detection for getting a conclusive result,
than the probability of correct detection attainable without
inconclusive results appearing \cite{CheB98,FJ03,Eld03}.

In this paper, we derive a new lower bound on the minimum-error
probability for ambiguous discrimination between arbitrary $m$
mixed quantum states with given prior probabilities. We show that
this bound improves, in some cases, the previous six lower bounds
in the literature, and also it betters the one derived in
\cite{Qiu08}. Also, we further present a necessary and sufficient
condition to show  how this new lower bound is attainable.

It is known that if $m=2$, the optimal inconclusive probability of
unambiguous discrimination $Q_{U}$ and the minimum-error
probability of ambiguous discrimination $Q_{E}$ have the
relationship $Q_{U}\geq 2Q_{E}$ \cite{HB04}. For $m>2$,   it was
proved in \cite{Qiu08} that $Q_{U}\geq 2Q_{E}$ holds only under
the restricted condition of the minimum-error probability
attaining the bound derived in \cite{Qiu08} (this restriction is
rigorous). In this paper, we show that, however, for $m>2$, the
relationship $Q_{U}\geq 2Q_{E}$ does not hold in general, which
may also reflect an essential difference between discrimination of
two-states and multi-states.

The remainder of the paper is organized as follows. In Section II,
we review six of the existing lower bounds on the minimum-error
probability for ambiguous discrimination between arbitrary $m$
mixed states and also give the new bound in this paper that will
be derived in the next section. Then, in Section III, we present
the new lower bound on the minimum-error probability for ambiguous
discrimination between arbitrary $m$ mixed states, and we give a
necessary and sufficient condition to show how this new lower
bound is attainable. Furthermore, in Section IV, we show that this
new bound improves the previous one in \cite{Qiu08}. In
particular, we try to compare these seven different lower bounds
reviewed in Section II with each other. Afterwards, in Section V,
we show that, for $m>2$, the relationship $Q_{U}\geq 2Q_{E}$ does
not hold in general, where $Q_{U}$ and $Q_{E}$ denote the optimal
inconclusive probability of unambiguous discrimination and the
minimum-error probability of ambiguous discrimination between
arbitrary $m$ mixed quantum states, respectively. Finally, some
concluding remarks are made in Section VI.

\section*{II. Reviewing the lower bounds on the minimum-error probability}

In this section, we review six of the existing lower bounds on
the minimum-error probability for ambiguous discrimination between
arbitrary $m$ mixed states. Also, we present the new bound in this
paper, but its proof is deferred to the next section.

Assume that a quantum system is described by a mixed quantum
state, say  $\rho$, drawn from a collection
$\{\rho_{1},\rho_{2},\ldots,\rho_{m}\}$ of mixed quantum states on
an $n$-dimensional complex Hilbert space ${\cal H}$, with the {\it
a priori} probabilities $p_{1},p_{2},\ldots,p_{m}$, respectively.
We assume without loss of generality that the all eigenvectors of
$\rho_{i}$, $1\leq i\leq m$, span ${\cal H}$, otherwise we
consider the spanned subspace instead of ${\cal H}$. A mixed
quantum state $\rho$ is a positive semidefinite operator with
trace 1, denoted $\textrm{Tr}(\rho)=1$. (Note that a positive
semidefinite operator must be a Hermitian operator
\cite{HJ86,NC00}.) To detect $\rho$, we need to design a
measurement consisting of $m$ positive semidefinite operators, say
$\Pi_{i}$, $1\leq i\leq m$, satisfying the resolution
\begin{equation}
\sum_{i=1}^{m}\Pi_{i}=I,
\end{equation}
where $I$ denotes the identity operator on ${\cal H}$.  By  the
measurement  $\Pi_{i}$, $1\leq i\leq m$, if the system has been
prepared by $\rho$, then $\textrm{Tr}(\rho\Pi_{i})$ is the
probability to deduce the system being state $\rho_{i}$.
Therefore, with this measurement the average probability $P$ of
correct detecting the system's state is as follows:
\begin{equation}
P=\sum_{i=1}^{m}p_{i}\textrm{Tr}(\rho_{i}\Pi_{i})
\end{equation}
and, the average probability $Q$ of erroneous detection is then as
\begin{equation}
Q=1-P=1-\sum_{i=1}^{m}p_{i}\textrm{Tr}(\rho_{i}\Pi_{i}).
\end{equation}
A main objective is to design an optimum measurement that
minimizes the probability of erroneous detection. As mentioned
above, for the case of $m=2$, the optimum detection problem has
been completely solved by Helstrom [4], and the minimum attainable
error probability, say $Q_{E}$, is by the Helstrom limit [4]
\begin{equation}
Q_{E}=\frac{1}{2}(1-\textrm{Tr}|p_{2}\rho_{2}-p_{1}\rho_{1}|),
\end{equation}
where $|A|=\sqrt{A^{\dag}A}$ for any linear operator $A$, and
$A^{\dag}$ denotes the conjugate transpose of $A$.

For discriminating more than two states, some bounds have been
obtained \cite{NS06,HKK06,Qiu08,Mon08,M07,T09,BK02,HLS05}, and we
review six \cite{NS06,HKK06,Qiu08,Mon08,M07,T09} of them in the
following. We first give a lower bound, and it follows from the
following lemma that is referred to \cite{NS06} by Nayak and
Salzman.

\begin{Lm}[\cite{NS06}] \label{r0}
If $0\leq\lambda_{i}\leq1$, and $\sum_{i=1}^{m}\lambda_{i}\leq l$,
then $\sum_{i=1}^{m}p_{i}\lambda_{i}\leq Pr(\{p_{i}\},l)$, where
$\{p_{1},p_{2},\ldots,p_{m}\}$ is a probability distribution, and
$Pr(\{p_{i}\},l)$ denotes the sum of the $l$ comparatively larger
probabilities of $\{p_{1},p_{2},\ldots,p_{m}\}$ (e.g., if
$p_{i_{1}}\geq p_{i_{2}}\geq\ldots\geq p_{i_{m}}$ and $l\leq m$,
then $Pr(\{p_{i}\},l)=\sum_{k=1}^{l}  p_{i_{k}}$).
\end{Lm}

From this lemma it follows a lower bound on the minimum-error
probability for ambiguous discrimination between
$\{\rho_{1},\rho_{2},\ldots,\rho_{m}\}$  with the {\it a priori}
probabilities $p_{1},p_{2},\ldots,p_{m}$. We first recall the
operator norm and trace norm of operator $A$.  $\|A\|$ denotes the
operator norm of $A$, i.e., $\|A\|=\max\{\|A|\phi\rangle\|:
|\psi\rangle \in {\cal S}\}$, where ${\cal S}$ is the set of all
unit vectors, that is to say, $\|A\|$
 is the largest singular value of $A$.
$\|A\|_{\textrm{tr}}=\textrm{Tr}\sqrt{A^{\dagger}A}$ denotes the
trace norm of $A$, equivalently, $\|A\|_{\textrm{tr}}$ is the sum
of the singular values of $A$.

\begin{Th} \label{L0}
For any $m$ mixed quantum states $\rho_{1},\ \rho_{2},\
\cdots,\ \rho_{m}$ with a $priori$ probabilities $p_{1},\ p_{2},\
\cdots,\ p_{m}$, respectively, then the minimum-error probability
$Q_{E}$ satisfies $Q_{E}\geq L_{0}$, where
\begin{eqnarray}
L_{0}=1-Pr(\{p_{i}\},d),
\end{eqnarray}
and $d$ denotes the dimension of the Hilbert space spanned by
$\{\rho_{i}\}$ .
\end{Th}

\begin{proof} Let $P_{S}$ denote the optimal correct probability, and let $\mathbb{E}_{m}$
denote the class of all POVM of the form $\{E_{i}:1\leq i\leq
m\}$. Due to
\begin{eqnarray}
\sum_{i=1}^{m}\textrm{Tr}(\rho_{i}E_{i})\leq\sum_{i=1}^{m}\|\rho_{i}\|\cdot\|E_{i}\|_{\textrm{tr}}
=\sum_{i=1}^{m}\|E_{i}\|_{\textrm{tr}}=\sum_{i=1}^{m}\textrm{Tr}(E_{i})=\textrm{Tr}(I)=d,
\end{eqnarray}
and with Lemma \ref{r0}, we have
\begin{eqnarray}
\sum_{i=1}^{m}p_{i}\textrm{Tr}(\rho_{i}E_{i})\leq Pr(\{p_{i}\},
d).
\end{eqnarray}
We get
\begin{eqnarray}
P_{S}=\max_{\{E_{j}\}\in\mathbb{E}_{m}}\sum_{i=1}^{m}p_{i}\textrm{Tr}(\rho_{i}E_{i})\leq
Pr(\{p_{i}\}, d),
\end{eqnarray}
Thus, we have
\begin{eqnarray}
Q_{E}=1-P_{S}\geq1-Pr\left(\{p_{i}\},d\right).
\end{eqnarray}
 The proof is completed.
\end{proof}

Another lower bound $L_{1}$ was given by Hayashi et al.
\cite{HKK06} in terms of the individual operator norm. That is,
\begin{eqnarray}
L_{1}=1-d\max_{i=1,\cdots,m}\{||p_{i}\rho_{i}||\},
\end{eqnarray}
where $d$, as above, is the dimension of the Hilbert space spanned
by $\{\rho_{i}\}$. It is easily seen that $L_{1}$ may be negative
for discriminating some states.

Recently, Qiu \cite{Qiu08} gave a lower bound $L_{2}$ in terms of
pairwise trace distance, i.e.,
\begin{eqnarray}
L_{2}=\frac{1}{2}\left(1-\frac{1}{m-1}\sum_{1\leq i<j\leq
m}\textrm{Tr}|p_{j}\rho_{j}-p_{i}\rho_{i}|\right).
\end{eqnarray}

Then, Montanaro \cite{Mon08} derived a lower bound $L_{3}$ in
terms of pairwise fidelity, that is,
\begin{eqnarray}
L_{3}=\sum_{1\leq i<j\leq m}p_{i}p_{j}F^{2}(\rho_{i},\rho_{j}),
\end{eqnarray}
where, also in this paper,
$F(\rho_{i},\rho_{j})=\textrm{Tr}\sqrt{\sqrt{\rho_{i}}\rho_{j}\sqrt{\rho_{i}}}$
as usual \cite{NC00}.

In this paper, we will derive a new lower bound $L_{4}$ in terms
of trace distance. More exactly,
\begin{eqnarray}
L_{4}=1-\min_{k=1,\cdots, m}\left(p_{k}+\sum_{j\neq
k}Tr(p_{j}\rho_{j}-p_{k}\rho_{k})_{+}\right),
\end{eqnarray}
where $(p_{j}\rho_{j}-p_{k}\rho_{k})_{+}$ denotes the positive
part of a spectral decomposition of $p_{j}\rho_{j}-p_{k}\rho_{k}$.
The proof for deriving $L_{4}$ is deferred to Section III.

Besides, Tyson \cite{T09} derived a lower bound $L_{5}$, that is,
\begin{eqnarray}
L_{5}=1-Tr\sqrt{\sum_{i=1}^{m}p_{i}^{2}\rho_{i}^{2}}.
\end{eqnarray}

Montanaro \cite{M07} derived a lower bound of pure states discrimination. For discriminating pure states $\{|\psi_{i}\rangle\}$ with a priori probabilities $p_{i}$, the minimum error probability satisfy
\begin{eqnarray}
Q_{E}^{*}\geq1-\sqrt{\sum_{i=1}^{m}(\langle\psi_{i}^{'}|\rho^{-\frac{1}{2}}|\psi_{i}^{'}\rangle)^{2}},
\end{eqnarray}
where $|\psi_{i}^{'}\rangle=\sqrt{p_{i}}|\psi_{i}\rangle$ and $\rho=\sum_{i=1}^{m}|\psi_{i}^{'}\rangle\langle\psi_{i}^{'}|$.
By the following lemma that is referred to Tyson \cite{T09}, a mixed state lower bound can be obtained from the pure-state lower bound.
\begin{Lm}[\cite{T09}]
Take spectral decompositions $\rho_{i}=\sum_{k}\lambda_{ik}|\psi_{ik}\rangle\langle\psi_{ik}|$, and consider the pure-state ensemble $\xi^{*}=\{(|\psi_{ik}\rangle, p_{i}\lambda_{ik})\}$. Then the minimum error probability $Q_{E}^{*}$ for discriminating $\xi^{*}$ satisfies
\begin{eqnarray}
Q_{E}\leq Q_{E}^{*}\leq(2-Q_{E})Q_{E}.
\end{eqnarray}
\end{Lm}

From the above lemma, we can get
\begin{eqnarray}
Q_{E}\geq1-\sqrt{1-Q_{E}^{*}}.
\end{eqnarray}
So, we get a lower bound for discriminating mixed state $\{\rho_{i}\}$, that is
\begin{eqnarray}
Q_{E}\geq1-\sqrt[4]{\sum_{i=1}^{m}\sum_{k=1}^{rank(\rho_{i})}(\langle\psi_{ik}^{'}|\rho^{-\frac{1}{2}}|\psi_{ik}^{'}\rangle)^{2}},
\end{eqnarray}
where $\rho=\sum_{i}^{m}p_{i}\rho_{i}$, $|\psi_{ik}^{'}\rangle=\sqrt{p_{i}\lambda_{ik}}|\psi_{ik}\rangle$, and $\rho_{i}=\sum_{k=1}^{rank(\rho_{i})}\lambda_{ik}|\psi_{ik}\rangle\langle\psi_{ik}|$. We denote this lower bound as
\begin{eqnarray}
L_{6}=1-\sqrt[4]{\sum_{i=1}^{m}\sum_{k=1}^{rank(\rho_{i})}(\langle\psi_{ik}^{'}|\rho^{-\frac{1}{2}}|\psi_{ik}^{'}\rangle)^{2}}.
\end{eqnarray}

\section*{III.  A new lower bound  and its attainability }

In this section, we derive the new lower bound $L_{4}$ on the
minimum-error discrimination between arbitrary $m$ mixed quantum
states, and then we give a sufficient and necessary condition to
achieve this bound.

The measures (e.g., various trace distances and fidelities)
between quantum states are of importance in quantum information
\cite{Uhl76,Joz94, FG99,NC00}. Here we first give three useful
lemmas concerning the usual trace distance and fidelity. As
indicated above, in this paper,
$F(\rho,\sigma)=\textrm{Tr}\sqrt{\sqrt{\rho}\sigma\sqrt{\rho}}$.

\begin{Lm}[\cite{NC00}]  \label{r00} Let $\rho$ and $\sigma$ be two quantum states. Then
\begin{eqnarray}
2(1-F(\rho,\sigma))\leq \textrm{Tr}|\rho-\sigma|\leq
2\sqrt{1-F^{2}(\rho,\sigma)}.
\end{eqnarray}

\end{Lm}

\begin{Lm}[\cite{Qiu08}] \label{r1} Let $\rho$ and $\sigma$ be two positive
semidefinite operators. Then
\begin{eqnarray}
\textrm{Tr}(\rho)+\textrm{Tr}(\sigma)-2F(\rho,\sigma)\leq
\textrm{Tr}|\rho-\sigma|\leq
\textrm{Tr}(\rho)+\textrm{Tr}(\sigma).
\end{eqnarray}
In addition, the second equality holds if and only if $\rho\bot
\sigma$.
\end{Lm}

\begin{Df}
Let $A$ be a self-adjoint matrix. Then the positive part is given
by
\begin{eqnarray}
A_{+}=\sum_{\lambda_{k}>0}\lambda_{k}\Pi_{k},
\end{eqnarray}
where $A=\sum_{k}\lambda_{k}\Pi_{k}$ is a spectral decomposition
of $A$.

\end{Df}

\begin{Lm} \label{r2}
Let $E$, $\rho$ and $\sigma$ are three positive semidefinite matrices, with
$E\leq I$. Then
\begin{eqnarray}
\textrm{Tr}(E(\rho-\sigma))\leq \textrm{Tr}(\rho-\sigma)_{+},
\label{TrE}
\end{eqnarray}
with equality iff $E$ is of the form
\begin{eqnarray}
E=P^{+}+P_{2},
\end{eqnarray}
where $P^{+}$ is the projection onto the support of
$(\rho-\sigma)_{+}$, and $0\leq P_{2}\leq I$ is supported on the
kernel of $(\rho-\sigma)$.
\end{Lm}

\begin{proof} See Appendix A.

\end{proof}

The new bound is presented by the following theorem.

\begin{Th} \label{r3} For any $m$ mixed quantum states $\rho_{1},\ \rho_{2},\
\cdots,\ \rho_{m}$ with a $priori$ probabilities $p_{1},\ p_{2},\\
\cdots,\ p_{m}$, respectively, then the minimum-error probability
$Q_{E}$ satisfies
\begin{eqnarray}
Q_{E}\geq L_{4}=1-\min_{k=1,\cdots, m}\left(p_{k}+\sum_{j\neq
k}Tr(p_{j}\rho_{j}-p_{k}\rho_{k})_{+}\right).\label{QE}
\end{eqnarray}
\end{Th}

\begin{proof} Let $P_{S}$ denote the maximum probability and let $\mathbb{E}_{m}$
denote the class of all POVM of the form $\{E_{i}:1\leq i\leq
m\}$. Then we have that, for any $k\in\{1,2,\ldots,m\}$,
\begin{eqnarray}
P_{S}&=&\max_{\{E_{j}\}\in\mathbb{E}_{m}} \sum_{j=1}^{m}\textrm{Tr}(E_{j}p_{j}\rho_{j})\\
&=&\max_{\{E_{j}\}\in\mathbb{E}_{m}}\left[p_{k}+\sum_{j\not=k}\textrm{Tr}(E_{j}(p_{j}\rho_{j}-p_{k}\rho_{k}))\right]\\
&\leq&p_{k}+\sum_{j\neq k}Tr(p_{j}\rho_{j}-p_{k}\rho_{k})_{+},
\label{th1.1}
\end{eqnarray}
where the inequality (\ref{th1.1}) holds by Lemma \ref{r2}.

Consequently, we get
\begin{eqnarray}
P_{S}\leq \min_{k=1,\cdots, m}\left(p_{k}+\sum_{j\neq
k}Tr(p_{j}\rho_{j}-p_{k}\rho_{k})_{+}\right).
\end{eqnarray}
Therefore, we conclude that inequality (\ref{QE}) holds by
$Q_{E}=1-P_{S}$.
\end{proof}

\begin{Rm} With Lemma \ref{r1},
$\textrm{Tr}|p_{j}\rho_{j}-p_{i}\rho_{i}|\leq p_{i}+p_{j} $, and
the equality holds if and only if $\rho_{j}\bot\rho_{i}$.
Therefore, in Theorem \ref{r3}, the upper bound on the probability
of correct detection between $m$ mixed quantum states satisfies
\begin{eqnarray}
&&p_{k_0}+\sum_{j\neq
k}Tr(p_{j}\rho_{j}-p_{k_0}\rho_{k_0})_{+}\nonumber\\
&=&\frac{1}{2}\left[1+\sum_{j\neq
k_{0}}\textrm{Tr}|p_{j}\rho_{j}-p_{k_{0}}\rho_{k_{0}}|-(m-2)p_{k_{0}}
\right]\nonumber\\
&\leq& \frac{1}{2}\left[1+\sum_{j\neq
k_{0}}(p_{j}+p_{k_{0}})-(m-2)p_{k_{0}} \right]=1.
\end{eqnarray}
By Lemma \ref{r1}, we further see that this bound is strictly
smaller than 1 usually unless $\rho_{1},\rho_{2},\cdots,\rho_{m}$
are mutually orthogonal.
\end{Rm}

\begin{Rm} When $m=2$, the lower bound in Theorem \ref{r3} is
precisely
$\frac{1}{2}(1-\textrm{Tr}|p_{2}\rho_{2}-p_{1}\rho_{1}|)$, which
is in accord with the well-known Helstrom limit \cite{Hel76}; and
indeed, in this case, this bound can always be attained by
choosing the optimum POVM: $E_{2}=P_{12}^{+}$ and $E_{1}=I-E_{2}$,
here $P_{12}^{+}$ denotes the projective operator onto the
subspace spanned by the all eigenvectors corresponding to all
positive eigenvalues of $p_{2}\rho_{2}-p_{1}\rho_{1}$. \end{Rm}

From the proof of Theorem \ref{r3}, we can obtain a sufficient and
necessary condition on the minimum-error probability $Q_{E}$
attaining the lower bound $L_{4}$, which is described by the
following theorem.

\begin{Th} \label{r4} Equality is attained in the bound (\ref{QE}) iff
for some fixed $k$, the operators
$\{(p_{j}\rho_{j}-p_{k}\rho_{k})_{+}\}_{j\neq k}$ have mutually
orthogonal supports.
\end{Th}

\begin{proof} See Appendix B.
\end{proof}

\section*{IV. Comparisons between the seven different lower bounds}

In this section, we compare the seven different lower bounds
($L_{i}, i=0,1,2,3,4,5,6$) on the minimum-error probability for
discriminating arbitrary $m$ mixed quantum states with the {\it a
priori} probabilities $p_{1},\ p_{2},\ \cdots,\ p_{m}$,
respectively. Also, when discriminating two states, we consider
their relation to Helstrom limit.

First, concerning the relation between $L_{4}$ and $L_{2}$, we
have the following result.

\begin{Th}\label{r5} For any $m$ mixed quantum states $\rho_{1},\ \rho_{2},\
\cdots,\ \rho_{m}$ with the {\it a priori} probabilities\\
$p_{1},\ p_{2},\ \cdots,\ p_{m}$, respectively, the two lower
bounds $L_{2}$ and $L_{4}$ on the minimum-error probability for
ambiguously discriminating these $m$ states have the following
relationship
\begin{eqnarray}
L_{4}\geq L_{2}.\label{AB}
\end{eqnarray}

\end{Th}

\begin{proof} First we recall
\begin{eqnarray} L_{4}&=&1-\min_{k=1,\cdots, m}\left(p_{k}+\sum_{j\neq
k}Tr(p_{j}\rho_{j}-p_{k}\rho_{k})_{+}\right)\\ &=&
\frac{1}{2}\left[1-\min_{k=1,\cdots, m}\left\{\sum_{j\neq
k}\textrm{Tr}|p_{j}\rho_{j}-p_{k}\rho_{k}|-(m-2)p_{k}
\right\}\right]
\end{eqnarray}
and
\begin{eqnarray}
L_{2}=\frac{1}{2}\left(1-\frac{1}{m-1}\sum_{1\leq i<j\leq
m}\textrm{Tr}|p_{j}\rho_{j}-p_{i}\rho_{i}|\right).
\end{eqnarray}
Let
\begin{eqnarray}
\min_{k=1,\cdots, m}\left\{\sum_{j\neq
k}\textrm{Tr}|p_{j}\rho_{j}-p_{k}\rho_{k}|-(m-2)p_{k} \right\}=
\sum_{j\neq
k_{0}}\textrm{Tr}|p_{j}\rho_{j}-p_{k_{0}}\rho_{k_{0}}|-(m-2)p_{k_{0}}
\label{th3.1}
\end{eqnarray}
for some $k_{0}\in \{1,2,\ldots,m\}$. Then
\begin{eqnarray}
L_{4}=\frac{1}{2}\left[1-\left(\sum_{j\neq
k_{0}}\textrm{Tr}|p_{j}\rho_{j}-p_{k_{0}}\rho_{k_{0}}|-(m-2)p_{k_{0}}\right)\right].
\end{eqnarray}
We can obtain the following inequality:
\begin{eqnarray}
2L_{4}-2L_{2}\geq\frac{m-2}{2(m-1)}-\frac{m-2}{2(m-1)}\left(\sum_{j\neq
k_{0}}\textrm{Tr}|p_{j}\rho_{j}-p_{k_{0}}\rho_{k_{0}}|-(m-2)p_{k_{0}}\right).\label{Ineq}
\end{eqnarray}
The proof of inequality (\ref{Ineq}) is arranged in Appendix C.

With Lemma \ref{r1}, we know that
$\textrm{Tr}|p_{j}\rho_{j}-p_{k_{0}}\rho_{k_{0}}|\leq
p_{j}+p_{k_{0}}$. Therefore, according to the inequality
(\ref{Ineq}), we further have
\begin{eqnarray}
& &2L_{4}-2L_{2} \nonumber\\
&\geq&\frac{m-2}{2(m-1)}-\frac{m-2}{2(m-1)}\left(\sum_{j\neq
k_{0}}(p_{j}+p_{k_{0}})-(m-2)p_{k_{0}}\right)\label{th3.3}\\
&=&\frac{m-2}{2(m-1)}-\frac{m-2}{2(m-1)}\left[1+(m-2)p_{k_{0}}-(m-2)p_{k_{0}}\right]\\
&=&0,
\end{eqnarray}
Consequently, we conclude that the inequality (\ref{AB}) holds and
the proof is completed.
\end{proof}

\begin{Emp} Indeed, $L_{4}>L_{2}$ is also possible for discriminating some states. Let $p_{1}=p_{2}=p_{3}=\frac{1}{3}$, and
$
\rho_{1}=\frac{1}{2}|0\rangle\langle0|+\frac{1}{2}|1\rangle\langle1|,
$ $
\rho_{2}=\frac{1}{3}|0\rangle\langle0|+\frac{2}{3}|2\rangle\langle2|,
$ $
\rho_{3}=\frac{1}{4}|0\rangle\langle0|+\frac{3}{4}|3\rangle\langle3|.
$ Then we can work out directly the seven lower bounds as: $
L_{0}=0, $ $ L_{1}=0, $ $ L_{2}=\frac{5}{36}, $ $
L_{3}=\frac{1}{24}, $ $ L_{4}=\frac{7}{36},$ $L_{5}=\frac{13-\sqrt{61}}{36}$ and $L_{6}=1-\sqrt[4]{\frac{10}{13}}$. Hence,
$L_{4}>L_{5}>L_{2}>L_{6}>L_{3}>L_{1}=L_{0}.$

Indeed,  the minimum-error probability $Q_{E}=\frac{7}{36}=L_{4}$.
We leave the calculation process out here, and we refer to the
method of calculation by using Lemma 13 in Section V.
\end{Emp}

In the sequel, we need another useful lemma.

\begin{Lm} \label{r6}
Let $\rho_{1}$ and $\rho_{2}$ be two mixed states, and
$p_{1}+p_{2}\leq 1$ with $p_{i}\geq 0$, $i=1,2$. Then
\begin{eqnarray}
p_{1}+p_{2}-2\sqrt{p_{1}p_{2}}F(\rho_{1},\rho_{2})\leq
\textrm{Tr}|p_{1}\rho_{1}-p_{2}\rho_{2}|\leq
p_{1}+p_{2}-2p_{1}p_{2}F^{2}(\rho_{1},\rho_{2}).
\end{eqnarray}
\end{Lm}

\begin{proof} Since $p_{1}\rho_{1}$ and $p_{2}\rho_{2}$ are positive semidefinite operators and
$F(p_{1}\rho_{1}, p_{2}\rho_{2})=\sqrt{p_{1}p_{2}}F(\rho_{1},
\rho_{2})$, we can directly get the first inequality from Lemma
\ref{r1}.

Now, we prove the second inequality. By Uhlmann's theorem
\cite{Uhl76,Joz94}, we let $|\psi_{1}\rangle$ and
$|\psi_{2}\rangle$ be the purifications of $\rho_{1}$ and
$\rho_{2}$, respectively, such that
$F(\rho_{1},\rho_{2})=|\langle\psi_{1}|\psi_{2}\rangle|$. Since
the trace distance is non-increasing under the partial trace
\cite{NC00}, we obtain
\begin{eqnarray}
\textrm{Tr}|p_{1}\rho_{1}-p_{2}\rho_{2}|\leq
\textrm{Tr}|p_{1}|\psi_{1}\rangle\langle\psi_{1}|-p_{2}|\psi_{2}\rangle\langle\psi_{2}||.
\end{eqnarray}
Let  $\{|\psi_{1}\rangle,|\psi_{1}^{\perp}\rangle\}$ be an
orthonormal basis in the subspace spanned by
$\{|\psi_{1}\rangle,|\psi_{2}\rangle\}$. Then $|\psi_{2}\rangle$
can be represented as
$|\psi_{2}\rangle=\cos\theta|\psi_{1}\rangle+\sin\theta|\psi_{1}^{\perp}\rangle$.
In addition, we have
\begin{eqnarray}
\textrm{Tr}|p_{1}|\psi_{1}\rangle\langle\psi_{1}|-p_{2}|\psi_{2}\rangle\langle\psi_{2}||=\textrm{Tr}\left|\left(%
\begin{array}{cc}
  p_{1}-p_{2}\cos^{2}\theta & -p_{2}\cos\theta \sin\theta \\
  -p_{2}\cos\theta \sin\theta & -p_{2}\sin^{2}\theta \\
\end{array}%
\right)\right|.
\end{eqnarray}
We can calculate the eigenvalues of the above matrix as
\begin{eqnarray}
\frac{1}{2}\left(p_{1}-p_{2}\pm\sqrt{p_{1}^{2}+p_{2}^{2}-2p_{1}p_{2}\cos(2\theta)}\right).
\end{eqnarray}
Therefore, we have
\begin{eqnarray}
\textrm{Tr}\left|p_{1}|\psi_{1}\rangle\langle\psi_{1}|-p_{2}|\psi_{2}\rangle\langle\psi_{2}|\right|=\sqrt{p_{1}^{2}+p_{2}^{2}-2p_{1}p_{2}\cos(2\theta)}.
\end{eqnarray}
Since
\begin{eqnarray}
2p_{1}p_{2}F^{2}(\rho_{1},\rho_{2})=2p_{1}p_{2}|\langle\psi_{1}|\psi_{2}\rangle|^{2}=2p_{1}p_{2}\cos^{2}\theta,
\end{eqnarray}
it suffices to show
\begin{eqnarray}
\sqrt{p_{1}^{2}+p_{2}^{2}-2p_{1}p_{2}\cos(2\theta)} \leq
p_{1}+p_{2}-2p_{1}p_{2}\cos^{2}\theta.
\end{eqnarray}
That is,
\begin{eqnarray}
p_{1}^{2}+p_{2}^{2}-2p_{1}p_{2}\cos(2\theta)\leq(p_{1}+p_{2}-2p_{1}p_{2}\cos^{2}\theta)^{2},
\end{eqnarray}
and equivalently,
\begin{eqnarray}
4p_{1}p_{2}\cos^{2}\theta[1-(p_{1}+p_{2})+p_{1}p_{2}\cos^{2}\theta]\geq
0,
\end{eqnarray}
which is clearly true. Consequently, we complete the proof.
\end{proof}

When $m=2$, we have the following relations between
$L_{2},L_{3},L_{4}$ and the Helstrom limit $H$.
\begin{Pp}\label{r7} When $m=2$,
\begin{eqnarray}
L_{4}=L_{2}=H\geq L_{3}, \label{2}
\end{eqnarray}
where $H$ is the Helstrom limit \cite{Hel76}, that is,
$H=\frac{1}{2}(1-\textrm{Tr}|p_{1}\rho_{1}-p_{2}\rho_{2}|)$.
\end{Pp}

\begin{proof}
 It is easy to verify that, when $m=2$,
$L_{1}=L_{2}=\frac{1}{2}(1-\textrm{Tr}|p_{1}\rho_{1}-p_{2}\rho_{2}|)=H$,
and $L_{3}=p_{1}p_{2}F^{2}(\rho_{1},\rho_{2})$. As a result, to
prove the inequality (\ref{2}), we should show that
$\frac{1}{2}(1-\textrm{Tr}|p_{1}\rho_{1}-p_{2}\rho_{2}|)\geq
p_{1}p_{2}F^{2}(\rho_{1},\rho_{2})$. Due to $p_{1}+p_{2}=1$,
according to the second inequality of Lemma \ref{r6}, we easily
get the conclusion, and therefore, (\ref{2}) holds.
\end{proof}

\begin{Rm}
From the proof of Lemma \ref{r6}, we know that when $m=2$, $L_{3}$
is smaller than Helstrom limit unless the mixed states are
mutually orthogonal.
\end{Rm}

Moreover,  if we discriminate $m$ equiprobable mixed states, i.e.,
the $m$ mixed states are chosen uniformly at random
($p_{i}=\frac{1}{m},i=1,2,\cdots,m$), then $L_{3}$ and $L_{4}$
have the following relationship.

\begin{Pp} \label{r8}
If $p_{i}=\frac{1}{m}\ (i=1,2\cdots,m)$, then we have $L_{4}\geq
L_{3}$.
\end{Pp}

\begin{proof} See Appendix D.

\end{proof}

Furthermore, even if the prior probabilities are not equal, under
some restricted conditions, $L_{2}$, $L_{3}$ and $L_{4}$ also have
certain relationships. We present a sufficient condition as
follows.

\begin{Pp} \label{r9} Let $a_{i}=\sum_{j\neq i}p_{i}p_{j}F^{2}(\rho_{i},\rho_{j})$.
Then $L_{2},L_{3}$ and $L_{4}$ have the following relationship:
for any $m\geq2$,
\begin{eqnarray}
L_{2}\geq\frac{1}{m-1}L_{3},
\end{eqnarray}
and when $\max_{i=1,\cdots,
m}\left\{a_{i}\right\}\geq\frac{1}{2}\sum_{i=1}^{m}a_{i}$, we have
\begin{eqnarray}
L_{4}\geq L_{3}.
\end{eqnarray}
\end{Pp}

\begin{proof} See Appendix E.

\end{proof}

\begin{Emp} $L_{0}=L_{1}>L_{4}>L_{3}>L_{6}>L_{2}>L_{5}$ is also possible. Let $p_{1}=p_{2}=p_{3}=\frac{1}{3}$, and
$\rho_{1}=|0\rangle\langle0|$, $ \rho_{2}=|+\rangle\langle+|, $ $
\rho_{3}=|1\rangle\langle1|, $ where
$|+\rangle=\frac{|0\rangle+|1\rangle}{\sqrt{2}}$. Then we can
calculate explicitly the values of the seven lower bounds: $
L_{0}=\frac{1}{3}, $ $ L_{1}=\frac{1}{3}, $ $
L_{2}=\frac{2-\sqrt{2}}{6}, $ $ L_{3}=\frac{1}{9}, $ $
L_{4}=\frac{2-\sqrt{2}}{3}, $ $
L_{5}=0, $ and $
L_{6}=1-\sqrt[4]{\frac{5+2\sqrt{2}}{12}}.$
\end{Emp}

\begin{Emp} $L_{1}=L_{4}>L_{2}>L_{5}>L_{6}>L_{3}>L_{0}$ is also possible. Let $p_{1}=p_{2}=p_{3}=\frac{1}{3}$, and
$
\rho_{1}=\frac{1}{2}|0\rangle\langle0|+\frac{1}{2}|1\rangle\langle1|,
$ $
\rho_{2}=\frac{1}{2}|0\rangle\langle0|+\frac{1}{2}|2\rangle\langle2|,
$ $
\rho_{3}=\frac{1}{2}|0\rangle\langle0|+\frac{1}{2}|3\rangle\langle3|.
$ Then we can calculate explicitly the values of the seven lower
bounds as: $ L_{0}=0, $ $ L_{1}=\frac{1}{3}, $ $
L_{2}=\frac{1}{4}, $ $ L_{3}=\frac{1}{12}, $ $
L_{4}=\frac{1}{3}, $ $L_{5}=\frac{3-\sqrt{3}}{6}$ and $L_{6}=1-\sqrt[4]{\frac{2}{3}}.$
\end{Emp}

\begin{Emp} $L_{4}>L_{5}>L_{3}>L_{2}>L_{6}>L_{0}>L_{1}$ is possible. Let $p_{1}=\frac{1}{10},\ p_{2}=\frac{1}{10},\ p_{3}=\frac{8}{10}$, and
$\rho_{1}=\frac{9}{10}|0\rangle\langle0|+\frac{1}{10}|1\rangle\langle1|,$
$\rho_{2}=\frac{9}{10}|0\rangle\langle0|+\frac{1}{10}|2\rangle\langle2|,$
$\rho_{3}=\frac{9}{10}|0\rangle\langle0|+\frac{1}{10}|3\rangle\langle3|.$
Similarly, we can calculate explicitly the values of the seven
lower bounds as: $ L_{0}=0, $ $ L_{1}=-\frac{47}{25}, $ $
L_{2}=\frac{1350}{10000}, $ $ L_{3}=\frac{1377}{10000}, $ $
L_{4}=\frac{1800}{10000}, $ $L_{5}=\frac{90-9\sqrt{66}}{100}$ and $L_{6}=1-\sqrt[4]{\frac{694}{1000}}.$
\end{Emp}

To sum up, when $m=2$, we have $L_{4}=L_{2}=H\geq L_{3}$ ($\geq$
can be strict for some states), and for any $m$ states, $L_{4}\geq
L_{2}$ always holds ($\geq$ can be strict for some states). For
the equiprobable case (the prior probabilities are equivalent),
$L_{4}\geq L_{3}$ always holds. Besides, in general, there are no
absolutely big and small relations between the other bounds, and
we have provided a number of examples to verify this result.

\section*{V. Comparison between ambiguous and unambiguous
discrimination}

For the sake of readability, we briefly recall the scheme of
unambiguous discrimination between mixed quantum states
$\{\rho_{i}: i=1 ,2, \cdots, m\}$ with the {\it a priori}
probabilities $\{p_{i}: i=1, 2, \cdots, m\}$, respectively. To
distinguish between $\rho_{i}$ unambiguously, we need to design a
measurement consisting of $m+1$ positive semidefinite operators,
say $\Pi_{i}$, $0\leq i\leq m$, satisfying the resolution
\begin{equation}
\sum_{i=0}^{m}\Pi_{i}=I,
\end{equation}
and, for $1\leq i,j\leq m$, if $i\not=j$,
\begin{equation}
\textrm{Tr}(\Pi_{i}\rho_{j})=0.
\end{equation}
$\Pi_{0}$ is related to the inconclusive result and $\Pi_{i}$
corresponds to an identification of $\rho_{i}$ for $1\leq i\leq
m$. Therefore, the average probability $P$ of correctly
distinguishing these  states is as follows:
\begin{equation}
P=\sum_{i=1}^{m}p_{i}\textrm{Tr}(\rho_{i}\Pi_{i})
\end{equation}
and, the average failure (inconclusive) probability $Q$  is then
as
\begin{equation}
Q=1-P=\sum_{i=1}^{m}p_{i}\textrm{Tr}(\rho_{i}\Pi_{0}).
\end{equation}

It is known that if $m=2$,  $Q_{U}$ and $Q_{E}$ have the
relationship $Q_{U}\geq2Q_{E}$ \cite{HB04}. For $m\geq 3$, it was
proved that, under the restricted condition of the minimum-error
probability attaining $L_{2}$, $Q_{U}\geq2Q_{E}$ still holds
\cite{Qiu08}.  A natural question  is that whether or not it still
holds  without any restricted condition. In this section, we will
prove that, however, for $m\geq3$, it may not hold again in
general. We can reuse the states of Example 1 to show this
conclusion.

\begin{Emp} Suppose that $\rho_{1},\rho_{2},\rho_{3}$ and
$p_{1},p_{2},p_{3}$ are the same as those in Example 1, that is,
$p_{1}=p_{2}=p_{3}=\frac{1}{3}$, and
$\rho_{1}=\frac{1}{2}|0\rangle\langle0|+\frac{1}{2}|1\rangle\langle1|$,
$\rho_{2}=\frac{1}{3}|0\rangle\langle0|+\frac{2}{3}|2\rangle\langle2|$,
$\rho_{3}=\frac{1}{4}|0\rangle\langle0|+\frac{3}{4}|3\rangle\langle3|$.
Then, for any POVM $\{E_{1},E_{2},E_{3}\}$,  by Lemma 4  we have
\begin{eqnarray}
&&\frac{1}{3}\times[\textrm{Tr} (E_{1}\rho_{1})+\textrm{Tr}(E_{2}\rho_{2})+\textrm{Tr}(E_{3}\rho_{3})]\nonumber\\
&=&\frac{1}{3}\times[1+\textrm{Tr}(E_{2}(\rho_{2}-\rho_{1}))+\textrm{Tr}(E_{3}(\rho_{3}-\rho_{1}))]\\
&\leq&\frac{1}{3}\times[1+\frac{1}{2}\textrm{Tr}|\rho_{2}-\rho_{1}|+\frac{1}{2}
\textrm{Tr}|\rho_{3}-\rho_{1}|\ ] \label{Emp1}\\
&=&\frac{29}{36}.
\end{eqnarray}
In particular, when $E_{2}=|2\rangle\langle2|$,
$E_{3}=|3\rangle\langle3|$, and $E_{1}=I-E_{2}-E_{3}$, the above
(\ref{Emp1}) becomes an equality. In other words, the average
success probability can achieve the upper bound $\frac{29}{36}$.
Therefore, we obtain the minimum-error probability $Q_{E}$ as
\begin{eqnarray}
Q_{E}=\frac{7}{36}, \label{Qe}
\end{eqnarray}
which, as calculated in Example 1, is equal to the lower bound
$L_{4}$, but not equal to the lower bound $L_{2}=\frac{5}{36}$.
(In the end of the section, we will recheck that
$Q_{E}=\frac{7}{36}$ holds exactly.)

Next, we consider the optimal inconclusive probability of
unambiguous discrimination $Q_{U}$. We have known that unambiguous
discrimination should satisfy the following two conditions:
\begin{eqnarray}
\textrm{Tr}(\Pi_{i}\rho_{j})=\delta_{ij}p_{i},\label{c1}\\
\Pi_{0}+\sum_{i=1}^{m}\Pi_{i}=I.
\end{eqnarray}
The condition (\ref{c1}) is also equivalent to
\begin{eqnarray}
\Pi_{i}\rho_{j}=0,
\end{eqnarray}
for $i\neq j,\ i,j=1,2,\cdots,m$.

As a result, in order to unambiguously discriminate the above
three states $\rho_{1},\rho_{2},\rho_{3}$, the POVM will be the
form: $\Pi_{1}=\alpha_{1}|1\rangle\langle1|$,
$\Pi_{2}=\alpha_{2}|2\rangle\langle2|$,
$\Pi_{3}=\alpha_{3}|3\rangle\langle3|$, and
$\Pi_{0}=I-\sum_{i=1}^{m}\Pi_{i}$, where $0\leq \alpha_{1},
\alpha_{2}, \alpha_{3}\leq 1$. Therefore,
\begin{eqnarray}
\frac{1}{3}\sum_{i=1}^{3}\textrm{Tr}(\Pi_{i}\rho_{i})=\frac{1}{3}\times[\frac{1}{2}\alpha_{1}+\frac{2}{3}\alpha_{2}+\frac{3}{4}\alpha_{3}]
\leq\frac{23}{36}.\label{Emp2}
\end{eqnarray}
When $\alpha_{1}=\alpha_{2}=\alpha_{3}=1$, the above (\ref{Emp2})
will be an equality. That is to say, the optimal success
probability can achieve this bound $\frac{23}{36}$. Therefore, we
have the optimal inconclusive probability $Q_{U}$ of unambiguous
discrimination between $\rho_{1},\rho_{2},\rho_{3}$ as follows:
\begin{eqnarray}
Q_{U}=\frac{13}{36}. \label{Qu}
\end{eqnarray}
Consequently, by combining (\ref{Qe}) and (\ref{Qu}) we have
\begin{eqnarray}
Q_{U}=\frac{13}{36}\not\geq2\times\frac{7}{36}=2Q_{E}.
\end{eqnarray}

\end{Emp}

To conclude, $Q_{U}\geq 2Q_{E}$ may not hold again if no condition
is imposed upon the discriminated states and prior probabilities.

A natural question is what is the supremum of $Q_U/Q_E$
for 3 states or $n$ states? Indeed, motivated by the above Example 5, we can give a more
general example to demonstrate that there is no supremum of $Q_{U}/Q_{E}$ for more than two states.

\begin{Emp}
 Assume that the three mixed states $\rho_{1},\
\rho_{2},\ \rho_{3}$ have the  {\it a priori} probabilities
$p_{1},\ p_{2},\ p_{3}$, respectively, where, for
$\alpha,\beta,\gamma\geq 0$, $
\rho_{1}=\alpha|0\rangle\langle0|+(1-\alpha)|1\rangle\langle1|, $
$ \rho_{2}=\beta|0\rangle\langle0|+(1-\beta)|2\rangle\langle2|, $
$ \rho_{3}=\gamma|0\rangle\langle0|+(1-\gamma)|3\rangle\langle3|.
$

First, we consider the optimal inconclusive probability of
unambiguous discrimination $Q_{U}$. Similar to Example 5, by
taking $\Pi_{1}=|1\rangle\langle1|$, $\Pi_{2}=|2\rangle\langle2|$,
$\Pi_{3}=|3\rangle\langle3|$, and
$\Pi_{0}=I-\sum_{i=1}^{m}\Pi_{i}=|0\rangle\langle 0|$, we can
obtain the optimal inconclusive probability $Q_{U}$ as
\begin{eqnarray}
Q_{U}=p_{1}\alpha+p_{2}\beta+p_{3}\gamma. \label{Emp3}
\end{eqnarray}
Then, we consider the minimum-error probability of ambiguous
discrimination $Q_{E}$. Note that
\begin{eqnarray}
p_{2}\rho_{2}-p_{1}\rho_{1}=(p_{2}\beta-p_{1}\alpha)|0\rangle\langle0|+p_{2}(1-\beta)|2\rangle\langle2|-p_{1}(1-\alpha)|1\rangle\langle1|,
\end{eqnarray}
\begin{eqnarray}
p_{3}\rho_{3}-p_{1}\rho_{1}=(p_{3}\gamma-p_{1}\alpha)|0\rangle\langle0|+p_{3}(1-\gamma)|3\rangle\langle3|-p_{1}(1-\alpha)|1\rangle\langle1|,
\end{eqnarray}
\begin{eqnarray}
p_{3}\rho_{3}-p_{2}\rho_{2}=(p_{3}\gamma-p_{2}\beta)|0\rangle\langle0|+p_{3}(1-\gamma)|3\rangle\langle3|-p_{2}(1-\beta)|2\rangle\langle2|.
\end{eqnarray}
If we let $p_{2}\beta\geq p_{1}\alpha\geq p_{3}\gamma$, then,
similar to Example 5, by taking $E_{1}=|1\rangle\langle1|$,
$E_{3}=|3\rangle\langle3|$, we can obtain
$E_{2}=I-E_{1}-E_{3}=|2\rangle\langle2|+|0\rangle\langle0|$, and
\begin{eqnarray}
Q_{E}=p_{1}\alpha+p_{3}\gamma.
\end{eqnarray}
Likewise, if $p_{1}\alpha\geq p_{2}\beta\geq p_{3}\gamma$, we can
get
\begin{eqnarray}
Q_{E}=p_{2}\beta+p_{3}\gamma,
\end{eqnarray}
and if $p_{3}\gamma\geq p_{1}\alpha\geq p_{2}\beta$, we have
\begin{eqnarray}
Q_{E}=p_{1}\alpha+p_{2}\beta.
\end{eqnarray}
In a word, we can always get that
\begin{eqnarray}
Q_{E}=p_{1}\alpha+p_{2}\beta+p_{3}\gamma-\max\left\{p_{1}\alpha,\
p_{2}\beta,\ p_{3}\gamma\right\}.
\end{eqnarray}
Consequently, with (\ref{Emp3}) we have
\begin{eqnarray}
Q_{U}/Q_{E}=\frac{p_{1}\alpha+p_{2}\beta+p_{3}\gamma}{p_{1}\alpha+p_{2}\beta+p_{3}\gamma-\max\left\{p_{1}\alpha,\
p_{2}\beta,\ p_{3}\gamma\right\}}.
\end{eqnarray}
Therefore, if we let $p_{1}\alpha=a$, $p_{2}\beta$ and $p_{3}\gamma$ be infinite small but not zero (As we know, this can be always preserved for appropriate $p_{i}$
($i=1,2,3$) and $\alpha,\beta,\gamma$), then $Q_{U}/Q_{E}$ will be infinite large. To conclude, there is no supremum of $Q_{U}/Q_{E}$ for more than two states.
\end{Emp}

\begin{Rm}
In fact, by virtue of a sufficient and necessary condition
regarding the minimum-error probability of ambiguous
discrimination, we can recheck the optimum measurement in Examples
5 and 6.
\end{Rm}

We recall this condition described by the following lemma,  that
is from \cite{Hel76,Hol73,YKL75,EMV03,BC08}.

\begin{Lm}[\cite{Hel76,Hol73,YKL75,EMV03,BC08}] \label{Lm}
$\{E_{i}:\ i=1,\cdots,m\}$ is an optimum measurement for achieving
the minimum-error probability of ambiguously discriminating the
mixed quantum states $\{\rho_{i}:\ i=1,\cdots,m\}$ with the a
priori probabilities $\{p_{i}:\ i=1,\cdots,m\}$, respectively, if
and only if
\begin{eqnarray}
R-p_{j}\rho_{j}\geq0,\ \ \forall j,
\end{eqnarray}
where the operator
\begin{eqnarray}
R=\sum_{i=1}^{m}p_{i}\rho_{i}E_{i}
\end{eqnarray}
is required to be Hermitian.
\end{Lm}

By utilizing Lemma \ref{Lm}, we can recheck the optimum
measurements in Examples 5.

In Example 5, by using the POVM
$E_{1}=|0\rangle\langle0|+|1\rangle\langle1|$,
$E_{2}=|2\rangle\langle2|$, $E_{3}=|3\rangle\langle3|$,  we obtain
that $Q_{E}=\frac{7}{36}$ is the minimum-error probability for
ambiguously discriminating $\rho_{1},\rho_{2},\rho_{3}$ with
$p_{1}=p_{2}=p_{3}=\frac{1}{3}$. Indeed, such a POVM is optimum by
Lemma \ref{Lm}. We can verify that
\begin{eqnarray}
\sum_{i=1}^{3}
p_{i}\rho_{i}E_{i}=\frac{1}{6}|0\rangle\langle0|+\frac{1}{6}|1\rangle\langle1|+\frac{2}{9}|2\rangle\langle2|+\frac{1}{4}|3\rangle\langle3|
\end{eqnarray}
is Hermitian, and
\begin{eqnarray}
\sum_{i=1}^{3}
p_{i}\rho_{i}E_{i}-p_{1}\rho_{1}=\frac{2}{9}|2\rangle\langle2|+\frac{1}{4}|3\rangle\langle3|\geq0,\\
\sum_{i=1}^{3}
p_{i}\rho_{i}E_{i}-p_{2}\rho_{2}=\frac{1}{18}|0\rangle\langle0|+\frac{1}{6}|1\rangle\langle1|+\frac{1}{4}|3\rangle\langle3|\geq0,\\
\sum_{i=1}^{3}
p_{i}\rho_{i}E_{i}-p_{3}\rho_{3}=\frac{1}{12}|0\rangle\langle0|+\frac{1}{6}|1\rangle\langle1|+\frac{2}{9}|2\rangle\langle2|\geq0.
\end{eqnarray}
By Lemma \ref{Lm}, we can conclude that
$E_{1}=|0\rangle\langle0|+|1\rangle\langle1|$,
$E_{2}=|2\rangle\langle2|$, $E_{3}=|3\rangle\langle3|$ compose an
optimum measurement. Therefore, we have the minimum error
probability $
Q_{E}=1-\sum_{i}Tr(p_{i}\rho_{i}E_{i})=1-(\frac{1}{3}+\frac{2}{9}+\frac{1}{4})=\frac{7}{36}.
$

In Example 6, we consider three cases:

1) If $\max\left(p_{1}\alpha,\ p_{2}\beta,\
p_{3}\gamma\right)=p_{1}\alpha$, then let
$E_{1}=|0\rangle\langle0|+|1\rangle\langle1|$,
$E_{2}=|2\rangle\langle2|$, $E_{3}=|3\rangle\langle3|$.

2) If $\max\left(p_{1}\alpha,\ p_{2}\beta,\
p_{3}\gamma\right)=p_{2}\beta$, then let
$E_{1}=|1\rangle\langle1|$,
$E_{2}=|0\rangle\langle0|+|2\rangle\langle2|$.
$E_{3}=|3\rangle\langle3|$.

3) If $\max\left(p_{1}\alpha,\ p_{2}\beta,\
p_{3}\gamma\right)=p_{3}\gamma$, then let
$E_{1}=|1\rangle\langle1|$, $E_{2}=|2\rangle\langle2|$,
$E_{3}=|0\rangle\langle0|+|3\rangle\langle3|$.

\noindent We can verify that
\begin{eqnarray}
\sum_{i=1}^{3} p_{i}\rho_{i}E_{i}=\max\left(p_{1}\alpha,\ p_{2}\beta,\
p_{3}\gamma\right)|0\rangle\langle0|+p_{1}(1-\alpha)|1\rangle\langle1|+p_{2}(1-\beta)|2\rangle\langle2|+p_{3}(1-\gamma)|3\rangle\langle3|\nonumber \end{eqnarray}
is Hermitian and, for each case,
\begin{eqnarray}
\sum_{i=1}^{3}p_{i}\rho_{i}E_{i}-p_{j}\rho_{j}\geq 0,  j=1,2,3.
\end{eqnarray}
By virtue of Lemma \ref{Lm}, we therefore obtain the
minimum-error probability $ Q_{E}=p_{1}\alpha+
p_{2}\beta+p_{3}\gamma-\max\left(p_{1}\alpha,\ p_{2}\beta,\
p_{3}\gamma\right). $

\section*{VI. Concluding remarks}

Quantum states discrimination is an intriguing issue in quantum
information processing
\cite{GRTZ02,Hel76,Che00,BHH04,Che04,BC08,EF01,Hol73}. In this
paper, we have reviewed a number of lower bounds on the
minimum-error probability for ambiguous discrimination between
arbitrary $m$ quantum mixed states. In particular, we have derived
a new lower bound on the minimum-error probability and presented a
sufficient and necessary condition for achieving this bound. Also,
we have proved that our bound improves the previous one obtained
in \cite{Qiu08}. In addition, we have compared the new bound with
six of the previous bounds, by a series of propositions and
examples. Finally, we have shown that, for $m>2$, the relationship
$Q_{U}\geq 2Q_{E}$ may not hold again in general, where $Q_{U}$
and $Q_{E}$ denote the optimal inconclusive probability of
unambiguous discrimination and the minimum-error probability of
ambiguous discrimination between arbitrary given $m$  mixed
quantum states, respectively. In addition,  we have demonstrated that there is no supremum of $Q_{U}/Q_{E}$ for more than two states by giving an
 example. As we know, for $m=2$, $Q_{U}\geq
2Q_{E}$ always holds \cite{HB04}, while for $m>2$, it holds only
under a certain restricted condition \cite{Qiu08}.

A further problem worthy of consideration is how to calculate the
minimum-error probability for ambiguous discrimination between
arbitrary $m$ quantum mixed states with the prior probabilities,
respectively, and devise an optimum measurement correspondingly.
In particular, we would consider the appropriate application of
these bounds presented in this paper in quantum communication
\cite{NS06,CP04}. Indeed, it is worth mentioning that quantum
state discrimination has already been applied to quantum encoding
\cite{EE07}.

\section*{Appendix A. The proof of Lemma \ref{r2}}

\begin{proof}
It is obvious that
\begin{eqnarray}
(\rho-\sigma)\leq (\rho-\sigma)_{+}
\end{eqnarray}
It follows immediately by the positivity of E (or by Lemma 2 of
Yuen-Kennedy-Lax \cite{YKL75}) that
\begin{eqnarray}
TrE(\rho-\sigma)\leq TrE(\rho-\sigma)_{+}.
\end{eqnarray}
Since $E\leq I$, it similarly follows that
\begin{eqnarray}
TrE(\rho-\sigma)_{+}\leq Tr(\rho-\sigma)_{+},
\end{eqnarray}
proving (\ref{TrE}). The equality condition is left as an exercise
for the reader.
\end{proof}

\section*{Appendix B. The proof of Theorem \ref{r4}}
\begin{proof}
Suppose for some $POVM$ $\{E_{k}\}$, we have equality in
\begin{eqnarray}
Tr\sum E_{k}\rho_{k}=Tr\left(\rho_{k}+\sum_{j\neq
k}E_{j}(\rho_{j}-\rho_{k})\right)\leq Tr\left(\rho_{k}+\sum_{j\neq
k}(\rho_{j}-\rho_{k})_{+}\right). \label{AppB1}
\end{eqnarray}
Then by Lemma 5
\begin{eqnarray}
E_{j}\geq \Pi_{+}(\rho_{j}-\rho_{k}),
\end{eqnarray}
where $\Pi_{+}(\rho_{j}-\rho_{k})$ is the positive projection onto
the positive subspace of $\rho_{j}-\rho_{k}$. If the unit vector
$|\psi\rangle$ is in the support of $(\rho_{j_{0}}-\rho_{k})_{+}$,
then one has
\begin{eqnarray}
1=|||\psi\rangle||^{2}=\sum_{j}\langle\psi|E_{j}|\psi\rangle=1+\sum_{j\neq
j_{0}}\langle\psi|E_{j}|\psi\rangle\geq 1.
\end{eqnarray}
It follows that $\langle\psi|E_{j}|\psi\rangle=0$ for all $j\neq
j_{0}$. In particular, the support of $E_{j_{0}}$ is orthogonal to
the supports of the other $E_{j}$'s.

Conversely, if the supports of the other $(\rho_{j}-\rho_{k})_{+}$
are mutually orthogonal, then the middle term of (\ref{AppB1})
attains a maximum for the $POVM$
\begin{eqnarray}
&&E_{j}=\Pi_{+}(\rho_{j}-\rho_{k}),\ j\neq k\\
&&E_{k}=I-\sum_{j\neq k}E_{j}.
\end{eqnarray}
In this case, one has equality of all terms in (\ref{AppB1}).
\end{proof}

\section*{Appendix C. The proof of inequality (\ref{Ineq})}
\begin{proof} First we recall that
\begin{eqnarray}
L_{4}&=&1-\min_{k=1,\cdots, m}\left(p_{k}+\sum_{j\neq
k}Tr(p_{j}\rho_{j}-p_{k}\rho_{k})_{+}\right)\\
&=&\frac{1}{2}\left[1-\left(\sum_{j\neq
k_{0}}\textrm{Tr}|p_{j}\rho_{j}-p_{k_{0}}\rho_{k_{0}}|-(m-2)p_{k_{0}}\right)\right],
\end{eqnarray}
and
\begin{equation}
L_{2}=\frac{1}{2}\left[1-\frac{1}{m-1}\sum_{1\leq i<j\leq
m}Tr|p_{j}\rho_{j}-p_{i}\rho_{i}|\right].
\end{equation}
Therefore,
\begin{eqnarray}
2L_{4}-2L_{2}=\frac{1}{m-1}\sum_{1\leq i<j\leq
m}\textrm{Tr}|p_{j}\rho_{j}-p_{i}\rho_{i}|-\left(\sum_{j\neq
k_{0}}\textrm{Tr}|p_{j}\rho_{j}-p_{k_{0}}\rho_{k_{0}}|-(m-2)p_{k_{0}}\right).\label{C1}
\end{eqnarray}
Note that
\begin{eqnarray}
\sum_{1\leq i<j\leq
m}\textrm{Tr}|p_{j}\rho_{j}-p_{i}\rho_{i}|=\frac{1}{2}\sum_{i=1}^{m}\sum_{j\neq
i}\textrm{Tr}|p_{j}\rho_{j}-p_{i}\rho_{i}|\label{C2}
\end{eqnarray}
and
\begin{eqnarray}
\sum_{j\neq
k_{0}}\textrm{Tr}|p_{j}\rho_{j}-p_{k_{0}}\rho_{k_{0}}|-(m-2)p_{k_{0}}=\frac{1}{m}\sum_{i=1}^{m}\left(\sum_{j\neq
k_{0}}\textrm{Tr}|p_{j}\rho_{j}-p_{k_{0}}\rho_{k_{0}}|-(m-2)p_{k_{0}}\right).\label{C3}
\end{eqnarray}
By combining Eqs. (\ref{C2},\ref{C3}) with Eq. (\ref{C1}), we have
\begin{eqnarray}
& &2L_{4}-2L_{2} \nonumber\\
&=&\frac{1}{2(m-1)}\sum_{i=1}^{m}\sum_{j\neq
i}\textrm{Tr}|p_{j}\rho_{j}-p_{i}\rho_{i}|-\frac{1}{m}\sum_{i=1}^{m}\left(\sum_{j\neq
k_{0}}\textrm{Tr}|p_{j}\rho_{j}-p_{k_{0}}\rho_{k_{0}}|-(m-2)p_{k_{0}}\right).\label{C4}
\end{eqnarray}
Furthermore, we can equivalently rewrite Eq. (\ref{C4}) as
follows:
\begin{eqnarray}
& &2L_{4}-2L_{2} \nonumber\\
&=&\frac{1}{2(m-1)}\sum_{i=1}^{m}\left[\left(\sum_{j\neq
i}\textrm{Tr}|p_{j}\rho_{j}-p_{i}\rho_{i}|-(m-2)p_{i}\right)
-\left(\sum_{j\neq
k_{0}}^{m}\textrm{Tr}|p_{j}\rho_{j}-p_{k_{0}}\rho_{k_{0}}|-(m-2)p_{k_{0}}\right)\right]\nonumber\\&&+\frac{m-2}{2(m-1)}
-(\frac{1}{m}-\frac{1}{2(m-1)})\sum_{i=1}^{m}\left(\sum_{j\neq
k_{0}}\textrm{Tr}|p_{j}\rho_{j}-p_{k_{0}}\rho_{k_{0}}|-(m-2)p_{k_{0}}\right)\label{C5}
\end{eqnarray}
With Eq. (\ref{th3.1}) we know that, for any
$i\in\{1,2,\ldots,m\}$,
\begin{eqnarray}
\sum_{j\neq
i}\textrm{Tr}|p_{j}\rho_{j}-p_{i}\rho_{i}|-(m-2)p_{i}\geq\sum_{j\neq
k_{0}}\textrm{Tr}|p_{j}\rho_{j}-p_{k_{0}}\rho_{k_{0}}|-(m-2)p_{k_{0}}.
\end{eqnarray}
Note that $\frac{1}{m}-\frac{1}{2(m-1)}=\frac{m-2}{2(m-1)}$.
Therefore, with Eq. (\ref{C5}) we have
\begin{eqnarray}
& &2L_{4}-2L_{2} \nonumber\\
&\geq&\frac{m-2}{2(m-1)}-\frac{m-2}{2m(m-1)}\sum_{i=1}^{m}\left(\sum_{j\neq
k_{0}}\textrm{Tr}|p_{j}\rho_{j}-p_{k_{0}}\rho_{k_{0}}|-(m-2)p_{k_{0}}\right)\label{th3.2}\\
&=&\frac{m-2}{2(m-1)}-\frac{m-2}{2(m-1)}\left(\sum_{j\neq
k_{0}}\textrm{Tr}|p_{j}\rho_{j}-p_{k_{0}}\rho_{k_{0}}|-(m-2)p_{k_{0}}\right)
\end{eqnarray}
which is the inequality (\ref{Ineq}) as desired.
\end{proof}

\section*{Appendix D. The proof of Proposition \ref{r8}}

\begin{proof}

If $p_{i}=\frac{1}{m}\ (i=1,2\cdots,m)$, we have
\begin{eqnarray}
L_{3}=\frac{1}{m^{2}}\sum_{i<j}F^{2}(\rho_{i},\rho_{j}),
\end{eqnarray}
and for any $k_{0}\in\{1,2,\cdots,m\}$,
\begin{eqnarray}
L_{4}&=&1-\min_{k=1,\cdots, m}\left(p_{k}+\sum_{j\neq
k}Tr(p_{j}\rho_{j}-p_{k}\rho_{k})_{+}\right)\\&=&
\frac{1}{2}\left[1-\min_{k=1,\cdots, m}\left\{\sum_{j\neq
k}\textrm{Tr}|p_{j}\rho_{j}-p_{k}\rho_{k}|-(m-2)p_{k}
\right\}\right]\\
&=& \frac{1}{2}\left[\frac{2m-2}{m}-\frac{1}{m}\min_{k=1,\cdots,
m}\left\{\sum_{j\neq k}\textrm{Tr}|\rho_{j}-\rho_{k}|
\right\}\right]\\
&\geq&\frac{1}{2}\left[\frac{2m-2}{m}-\frac{1}{m}\sum_{j\neq
k_{0}}\textrm{Tr}|\rho_{j}-\rho_{k_{0}}|\right]\\
&\geq&\frac{1}{2}\left[\frac{2m-2}{m}-\frac{2}{m}\sum_{j\neq
k_{0}}\sqrt{1-F^{2}(\rho_{j},\rho_{k_{0}})}\right],
\end{eqnarray}
where the last inequality holds by Lemma \ref{r00}. Thus, we get
\begin{eqnarray}
L_{4}&\geq&\frac{1}{2}\left[\frac{2m-2}{m}-\frac{2}{m}\min_{k=1,\cdots,
m}\left\{\sum_{j\neq
k}\sqrt{1-F^{2}(\rho_{j},\rho_{k})}\right\}\right].\label{P6.1}
\end{eqnarray}
Therefore, we have
\begin{eqnarray}
&&2m^{2}(L_{4}-L_{3})\nonumber\\
&\geq& 2m^{2}-2m-2m\min_{k=1,\cdots,
m}\left\{\sum_{j\neq
k}\sqrt{1-F^{2}(\rho_{j},\rho_{k})}\right\}-2\sum_{i<j}F^{2}(\rho_{i},\rho_{j})\\
&=& 2m^{2}-2m-2\sum_{i=1}^{m}\min_{k=1,\cdots,
m}\left\{\sum_{j\neq
k}\sqrt{1-F^{2}(\rho_{j},\rho_{k})}\right\}-\sum_{i=1}^{m}\sum_{j\neq i}F^{2}(\rho_{i},\rho_{j})\\
&\geq&2m^{2}-2m-2\sum_{i=1}^{m}\sum_{j\neq
i}\sqrt{1-F^{2}(\rho_{j},\rho_{i})}-\sum_{i=1}^{m}\sum_{j\neq i}F^{2}(\rho_{i},\rho_{j})\label{P6.2}\\
&=&\sum_{i=1}^{m}\sum_{j\neq i}\left(\sqrt{1-F^{2}(\rho_{j},\rho_{i})}-1\right)^{2}\\
&\geq&0.
\end{eqnarray}
Thus, we have $L_{4}\geq L_{3}$. We complete the proof.
\end{proof}

\section*{Appendix E. The proof of Proposition \ref{r9}}

\begin{proof} By Lemma \ref{r6}, we have
\begin{eqnarray}
L_{2}&=&\frac{1}{2}\left(1-\frac{1}{m-1}\sum_{1\leq i<j\leq
m}\textrm{Tr}|p_{j}\rho_{j}-p_{i}\rho_{i}|\right)\\
&\geq&\frac{1}{2}\left(1-\frac{1}{m-1}\sum_{1\leq i<j\leq
m}[p_{i}+p_{j}-2p_{i}p_{j}F^{2}(\rho_{i},\rho_{j})]\right)\\
&=&\frac{1}{m-1}L_{3}.
\end{eqnarray}
For any given $k_{0}=1,\cdots,m$,
\begin{eqnarray}
L_{4}&=&1-\min_{k=1,\cdots, m}\left(p_{k}+\sum_{j\neq
k}Tr(p_{j}\rho_{j}-p_{k}\rho_{k})_{+}\right)\\&=&
\frac{1}{2}\left[1-\min_{k=1,\cdots, m}\left\{\sum_{j\neq
k}\textrm{Tr}|p_{j}\rho_{j}-p_{k}\rho_{k}|-(m-2)p_{k}
\right\}\right]\\
&\geq& \frac{1}{2}\left[1-\left(\sum_{j\neq
k_{0}}\textrm{Tr}|p_{j}\rho_{j}-p_{k_{0}}\rho_{k_{0}}|-(m-2)p_{k_{0}}
\right)\right]\\
&=&\frac{1}{2}-\frac{1}{2}\sum_{j\neq
k_{0}}\textrm{Tr}|p_{j}\rho_{j}-p_{k_{0}}\rho_{k_{0}}|+\frac{m-2}{2}p_{k_{0}}\\
&\geq&\frac{1}{2}-\frac{1}{2}\sum_{j\neq
k_{0}}[p_{k_{0}}+p_{j}-2p_{k_{0}}p_{j}F^{2}(\rho_{k_{0}},\rho_{j})]+\frac{m-2}{2}p_{k_{0}}\\
&=&\sum_{j\neq k_{0}}p_{k_{0}}p_{j}F^{2}(\rho_{k_{0}},\rho_{j}).
\end{eqnarray}
So, we have
\begin{eqnarray}
L_{4}\geq\max_{k=1,\cdots, m}\left\{\sum_{j\neq
k}p_{k}p_{j}F^{2}(\rho_{k},\rho_{j})\right\}.
\end{eqnarray}
Moreover, we have
\begin{eqnarray}
L_{3}=\sum_{1\leq i<j\leq
m}p_{i}p_{j}F^{2}(\rho_{i},\rho_{j})=\frac{1}{2}\sum_{i=1}^{m}\sum_{j\neq
i}p_{i}p_{j}F^{2}(\rho_{i},\rho_{j}).
\end{eqnarray}
Let $a_{i}=\sum_{j\neq i}p_{i}p_{j}F^{2}(\rho_{i},\rho_{j})$. Then
we get
\begin{eqnarray}
L_{4}-L_{3}\geq\max_{i=1,\cdots,
m}\left\{a_{i}\right\}-\frac{1}{2}\sum_{i=1}^{m}a_{i}.
\end{eqnarray}
If $\max_{i=1,\cdots,
m}\left\{a_{i}\right\}-\frac{1}{2}\sum_{i=1}^{m}a_{i}\geq0$, then
$L_{4}\geq L_{3}$. We complete the proof.
\end{proof}


\begin{thebibliography}{AB}


\bibitem{GRTZ02} N. Gisin, G. G. Ribordy, W. Tittel, and H. Zbinden, ``Quantum cryptography," \emph{Reviews of Modern Physics}, vol. 74, no.1,
 pp. 145-195, 2002.
\bibitem{Hel76} C. W. Helstrom, \emph{Quantum Detection and Estimation
Theory}, New York: Academic Press, 1976.
\bibitem{Che00} A. Chefles, ``Quantum state discrimination," \emph{Contemp.  Phys.}, vol. 41, pp. 401-424, 2000.
\bibitem{BHH04} J. A. Bergou, U. Herzog, and M. Hillery, ``Discrimination of Quantum
States,"\emph{Quantum State Estimation, Lecture Notes in Physics},
Vol. 649, pp. 417-465, Berlin: Springer, 2004.
\bibitem{Che04} A. Chefles, ``Quantum States: Discrimination and Classical
Information Transmission. A Review of Experimental Progress,"
\emph{Quantum State Estimation, Lecture Notes in Physics}, Vol.
649, pp. 467-511, Berlin: Springer, 2004.
\bibitem{BC08} S. M. Barnett and S. Croke, ``Quantum state discrimination," \emph{Advances in Optics and Photonics}, vol. 1, no. 2, pp.
238-278, Apr. 2009.
\bibitem{EF01} Y. C. Eldar and G. D. Forney, ``On quantum detection and the square-root measurement," \emph{IEEE Trans. Inf. Theory}, vol. 47, pp.
858-872, Mar. 2001.
\bibitem{Hol73}A.S. Holevo, ``Statistical decision theory for quantum systems," \emph{J. Multivariate Anal.}, vol. 3,
pp. 337-394, Dec. 1973.
\bibitem{YKL75}H. P. Yuen, R. S. Kennedy, and M. Lax, ``Optimum testing of multiple hypotheses in quantum detection theory," \emph{IEEE Trans. Inf. Theory}, vol. 21, pp. 125-134, Mar 1975.
\bibitem{Qiu08} D. W. Qiu, ``Minimum-error discrimination between mixed quantum states," \emph{Phys. Rev. A}, vol. 77, pp. 012328-012328, 2008.

\bibitem{BR97}S.M. Barnett and E. Riis, ``Experimental demonstration of
polarization discrimination at the Helstrom bound," \emph{J. Mod.
Opt.}, vol. 44, pp. 1061-1064, 1997.
\bibitem{CKCBR01}R. B. M. Clarke, V. M. Kendon, A. Chefles, S. M.
Barnett, and E. Riis, ``Experimental realization of optimal
detection strategies for overcomplete states," \emph{Phys. Rev.
A}, vol. 64, pp. 012303-012303, 2001.
\bibitem{MSB04}M. Mohseni, A. M. Steinberg, and J.A. Bergou, ``Optical Realization of Optimal Unambiguous Discrimination for Pure and Mixed Quantum States," \emph{Phys. Rev. Lett.}, vol. 93, pp. 200403-200403, 2004.
\bibitem{CBH89}M. Charbit, C. Bendjaballah, and C. W. Helstrom, ``Cutoff rate for the M-ary PSK modulation channel withoptimal quantum detection," \emph{IEEE Trans. Inf. Theory}, vol. 35, pp. 1131-1133, Sep. 1989.
\bibitem{EMV03} Y. C. Eldar, A. Megretski, and G.C. Verghese, ``Designing optimal quantum detectors via semidefinite programming," \emph{IEEE Trans. Inf. Theory}, vol. 49, 1007-1012, April 2003.
\bibitem{OBH96}M. Osaki, M. Ban, and O. Hirota, ``Derivation and physical interpretation of the optimum detection operators for coherent-state signals," \emph{Phys. Rev.
A}, vol. 54, pp. 1691-1691, 1996.
\bibitem{BKMH97} M. Ban, K. Kurokawa, R. Momose, and O. Hirota,``Optimum measurements for discrimination among symmetric quantum states and parameter estimation,"  \emph{Int. J. Theor. Phys.},
vol. 36, pp. 1269-1288, 1997.
\bibitem{EMV04}Y. C. Eldar, A. Megretski, and G. C. Verghese, ``Optimal Detection of Symmetric Mixed Quantum States,"
 \emph{IEEE Trans. Inf. Theory}, vol. 50, pp. 1198-1207, Jun. 2004.
\bibitem{Bar01} S.M. Barnett, ``Minimum-error discrimination between multiply symmetric states," \emph{Phys. Rev. A}, vol. 64, pp. 030303-030303, 2001.
\bibitem{ABGH02} E. Andersson, S.M. Barnett, C.R. Gilson, and K. Hunter, ``Minimum-error discrimination between three mirror-symmetric states," \emph{Phys. Rev. A}, vol. 65, pp. 052308, 2002.

\bibitem{CH03} C.-L. Chou and L. Y. Hsu, ``Minimum-error discrimination between symmetric mixed quantum states," \emph{Phys. Rev.
A}, vol. 68, pp. 042305-042305, 2003.
\bibitem{HKK06} M. Hayashi, A. Kawachi, and H. Kobayashi, ``Quantum measurements for hidden subgroup problems with optimal sample," \emph{Quantum Information and Computation,} vol. 8, pp. 0345-0358, 2008.
\bibitem{Mon08} A. Montanaro, ``A lower bound on the probability of error in quantum state discrimination," \emph{IEEE Inf. Theory Workshop}, pp. 378-380, May 2008.
\bibitem{NS06} A. Nayak, J. Salzman, ``Limits on the ability of quantum states to convey classical messages," \emph{Journal of the ACM}, vol. 53, no. 1, pp. 184-206, Jan.
2006.
\bibitem{M07} A. Montanaro, ``On the distinguishability of random quantum states," \emph{Communications in Mathematical Physics}, vol. 273,
no. 3, pp. 619-636, 2007.
\bibitem{T09} J. Tyson, ``Two-sided estimates of minimum-error distinguishability of mixed quantum states via generalized Holevo-Curlander bounds," \emph{J. Math Phys}, vol. 50, no. 3, pp.
032106-032106, 2009.
\bibitem{BK02} H. Barnum and E. Knill, ``Reversing quantum dynamics with near-optimal quantum and classical fidelity," {\it J. Math. Phys.}, vol. 43, no. 5, pp. 2097-2097, 2002.
\bibitem{HLS05} P. Hayden, D. Leung and G. Smith, ``Multiparty data hiding of quantum information," \emph{Phys. Rev. A}, vol. 71, pp. 062339, 2005.
\bibitem{Iva87} I. D. Ivanovic, ``How to differentiate between non-orthogonal states," \emph{Phys. Lett. A}, vol. 123, pp. 257-259, 1987.
\bibitem{Die88} D. Dieks, ``Overlap and distinguishability of quantum states," \emph{Phys. Lett. A}, vol. 126, pp. 303-306, 1988.

\bibitem{Per88} A. Peres, ``How to differentiate between non-orthogonal states," \emph{Phys. Lett. A}, vol. 128, pp. 19-19, 1988.
\bibitem{JS95} G. Jaeger and A. Shimony, ``Optimal distinction between two non-orthogonal quantum states,"  \emph{Phys. Lett. A}, vol. 197, pp. 83-87, Jan. 1995.
\bibitem{PT98}A. Peres and D. R. Terno,``Optimal distinction between
non-orthogonal quantum states," \emph{J. Phys. A: Math. Gen}, vol.
31, pp. 7105-7111, 1998.
\bibitem{DG98}L.M. Duan and G.C. Guo, ``Probabilistic Cloning and Identification of Linearly Independent Quantum States," \emph{Phys. Rev. Lett.}, vol. 80, pp.
4999-5002, 1998.
\bibitem{Che98}A. Chefles, ``Unambiguous discrimination between linearly independent quantum states," \emph{Phys. Lett. A}, vol. 239, pp. 339-347, 1998.
\bibitem{CB98} A. Chefles, S.M. Barnett, ``Optimum unambiguous discrimination between linearly independent symmetric states," \emph{Phys. Lett. A}, vol. 250, pp. 223-229, 1998.
\bibitem{Eld03IT} Y.C. Eldar, ``A semidefinite programming approach to optimal unambiguous discrimination of quantum states," \emph{IEEE Trans. Inf. Theory}, vol 49, pp.
446-456, Feb. 2003.
\bibitem{Qiu02L}  D.W. Qiu, ``Upper bound on the success probability for unambiguous discrimination," \emph{ Phy. Lett. A}, vol. 303, pp. 140-146, Oct. 2002.
\bibitem{Qiu02J}  D.W. Qiu, ``Upper bound on the success probability of separation among quantum states," \emph{J. Phys. A: Math. Gen.}, vol. 35, pp. 6931-6937, 2002.
\bibitem{Zha01} S. Zhang, Y. Feng, X. Sun, and M. Ying, ``Upper bound for the success probability of unambiguous discrimination among quantum states," \emph{Phys. Rev. A}, vol. 64, pp.
062103-062103, 2001.

\bibitem{RST03} T. Rudolph, R.W. Spekkens, and P.S. Turner, ``Unambiguous discrimination of mixed states," \emph{Phys. Rev.
A}, vol. 68, pp. 010301-010301, 2003.
\bibitem{FDY04} Y. Feng, R. Y. Duan, and M. Ying, ``Unambiguous discrimination between mixed quantum states," \emph{Phys. Rev. A}, vol. 70,
pp. 012308-012308, 2004.
\bibitem{RLE03}P. Raynal, N. L\"{u}tkenhaus, and S. J. van Enk, ``Reduction theorems for optimal unambiguous state discrimination of density matrices," \emph{Phys. Rev.
A}, vol. 68, pp. 022308-022308, 2003.
\bibitem{HB05} U. Herzog and J.A.  Bergou, ``Optimum unambiguous discrimination of two mixed quantum states," \emph{Phys. Rev. A}, vol. 71, pp.
050301-050301, 2005.
\bibitem{CheB98} A. Chefles and S.M. Barnett, ``Strategies for discriminating between non-orthogonal quantum states," \emph{J. Mod. Opt.}, vol. 45, pp. 1295-1302, 1998.
\bibitem{FJ03} J. Fiur\'{a}\v{s}ek, M. Je\v{z}ek, ``Optimal discrimination of mixed quantum states involving inconclusive results," \emph{Phys. Rev.
A}, vol. 67, pp. 012321-012321, 2003.
\bibitem{Eld03} Y.C. Eldar, ``Mixed-quantum-state detection with inconclusive results," \emph{Phys. Rev.
A}, vol. 67, pp. 042309-042309, 2003.
\bibitem{HB04}U. Herzog and J. A. Bergou, ``Distinguishing mixed quantum states: Minimum-error discrimination versus optimum unambiguous discrimination," \emph{Phys. Rev. A}, vol. 70, pp. 022302-022302, 2004.
\bibitem{HJ86} R.A. Horn, C.R. Johnson, \emph{Matrix Analysis},
Cambridge: Cambridge University Press, 1986.
\bibitem{NC00}M.A. Nielsen and I.L. Chuang, \emph{Quantum Computation and Quantum
Information}, Cambridge: Cambridge University Press, 2000.

\bibitem{Uhl76} A. Uhlmann, ``The `transition probability' in the
state space of a*-algebra," {\it Reps. Math. Phys.}, vol. 9, pp.
273-279, 1976.
\bibitem{Joz94} R. Jozsa, ``Fidelity for mixed quantum states," {\it J. Modern. Opt.}, vol. 41, no. 12, pp.
2315-2323, 1994.
\bibitem{FG99} C. A. Fuchs and J. van de Graaf, ``Cryptographic Distinguishability Measures for Quantum Mechanical States," {\it IEEE Trans. Inf. Theory}, vol.
45, pp. 1216-1227, 1999.
\bibitem{CP04} J. I. Concha and H. V. Poor, ``Multiaccess quantum channels," {\it IEEE Trans. Inf. Theory},
vol. 50, no. 5, pp. 725-747, May 2004.
\bibitem{EE07} N. Elron and Y. C. Eldar, ``Optimal Encoding of Classical
Information in a Quantum Medium," {\it IEEE Trans. Inf. Theory},
vol. 53, no. 5, pp. 1900-1907, May 2007.


\end{thebibliography}
\end{document}